\documentclass[aps,prd,nofootinbib,showpacs,twocolumn,floatfix]{revtex4}
\usepackage{amssymb,amsfonts}
\def \bea {\begin{eqnarray}}
\def \eea {\end{eqnarray}}

\begin{document}

\title{Inner boundary conditions for black hole Initial Data derived from
Isolated Horizons}

\author{Jos\'e Luis Jaramillo}
\email[]{jose-luis.jaramillo@obspm.fr} \affiliation{Laboratoire de
l'Univers et de ses Th\'eories, UMR 8102 du C.N.R.S., Observatoire
de Paris, F-92195 Meudon Cedex, France}

\author{Eric Gourgoulhon}
\email[]{eric.gourgoulhon@obspm.fr} \affiliation{Laboratoire de
l'Univers et de ses Th\'eories, UMR 8102 du C.N.R.S., Observatoire
de Paris, F-92195 Meudon Cedex, France}

\author{Guillermo A. Mena Marug\'an}
\email[]{mena@iem.cfmac.csic.es} \affiliation{Instituto de
Estructura de la Materia, Centro de F\'{\i}sica Miguel A. Catal\'an,
C.S.I.C., Serrano 121, 28006 Madrid, Spain}


\begin{abstract}
We present a set of boundary conditions for solving the elliptic
equations in the Initial Data Problem for space-times containing a
black hole, together with a number of constraints to be satisfied
by the otherwise freely specifiable standard parameters of the
Conformal Thin Sandwich formulation. These conditions altogether
are sufficient for the construction of a horizon that is {\it
instantaneously} in equilibrium in the sense of the Isolated
Horizons formalism. We then investigate the application of these
conditions to the Initial Data Problem of binary black holes and
discuss the relation of our analysis with other proposals that
exist in the literature.
\end{abstract}


\maketitle

\section{Introduction} \label{s:intro}

The problem of determining appropriate initial data for binary
black holes is of crucial importance in order to construct
successful numerical simulations for these astrophysical systems
\cite{Co00}. Starting with Einstein field equations, a specific
strategy for this problem consists in solving the relevant
elliptic equations on an initial Cauchy surface where a sphere
${\cal S}$ has been excised \cite{Th87} for each black hole (inner
boundary). The purpose of the present work is to present a set of
{\it inner} boundary conditions inspired by purely geometrical
considerations, inasmuch as they are derived from the formalism of
Isolated Horizons \cite{iso1,iso2,iso3,AFK00,ABL01a,ABL01b}, and
which guarantee that the excised sphere is in fact a section of a
quasi-equilibrium horizon.

A pioneering work on this inner boundary problem was presented by
Cook in Ref. \cite{Co01}, in the context of formulating a definite
full prescription for the construction of initial data for binary
black holes in quasi-circular orbits. The assumptions made in that
analysis permit to determine a proper set of conditions for a
quasi-equilibrium black hole. However, at least intuitively, one
would expect that the Isolated Horizons formalism, which is mainly
a systematic characterization of the notion and properties of
quasi-equilibrium horizons, could supply a more powerful and
consistent framework for discussing the conditions in this black
hole regime. Actually, the spirit in Ref. \cite{Co01} closely
resembles that encoded in the Isolated Horizons scheme, but does
not fully capture it. Therefore, in this specific sense, the
quasi-equilibrium horizon analysis may be refined. With this
motivation, we will truly adopt here the Isolated Horizons
formalism as the guideline of a geometrical analysis whose
ultimate goal is the {\it ab initio} numerical construction of an
isolated horizon. This strategy provides us with a rigorous
mathematical and conceptual framework that systematizes the
physical assumptions.

For the sake of clarity, we have considered important to provide a
relatively self-contained presentation, even at the cost of
lengthening the article. The rest of the work is organized as
follows. As in Refs. \cite{Co01,GGB02}, we use a Conformal Thin
Sandwich approach to set the Initial Data Problem; thus Sec. II
briefly reviews the basics of this approach. Sec. III introduces
the main ideas of the Isolated Horizons framework and underlines
the importance of its hierarchical structure by first introducing
Non-Expanding Horizons and then Weakly Isolated Horizons. Boundary
conditions on the horizon are derived in Sec. IV. Sec. V discusses
the relation of this approach with that of Ref. \cite{Co01}.
Finally Sec. VI presents the conclusions.

\section{Conformal Thin Sandwich approach to Initial Data}
\label{s:CFS}

In this section we formulate the problem that will be analyzed in
this work and introduce our notation. We will use Greek letters
$(\mu,\nu,...)$ for Lorentzian indices, intermediate Latin letters
$(i, j, ...)$ for spatial indices on a Cauchy slice, and Latin
letters from the beginning of the alphabet $(a, b, ...)$ for
coordinates on a two-dimensional sphere ${S^2}$.

Adopting a standard 3+1 decomposition for General Relativity (see
e.g. Ref. \cite{York79}), the space-time ${\cal M}$ with
Lorentzian metric $g_{\mu\nu}$ is foliated by spacelike
hypersurfaces $\Sigma_t$ parametrized by a scalar function $t$.
The evolution vector $t^\mu$, normalized as $t^\mu\nabla_\mu t=1$,
is decomposed in its normal and tangential parts by introducing
the lapse function $\alpha$ and the shift vector $\beta^\mu$ \bea
t^\mu=\alpha n^\mu + \beta^\mu \ , \label{alphabeta} \eea where
$n_\mu = -\alpha\nabla_\mu t$ is the unit timelike vector normal
to $\Sigma_t$ and $n^\mu \beta_\mu=0$.

Denoting by $\gamma_{ij}$ the induced metric on $\Sigma_t$, the
Lo\-rentz\-ian line element reads \begin{eqnarray} ds^2 &=&
g_{\mu\nu}dx^\mu dx^\nu \nonumber\\ &=& -\alpha^2 dt^2 +
\gamma_{ij}(dx^i + \beta^i dt) (dx^j + \beta^j dt) \ .
\end{eqnarray} The embedding of the hypersurfaces $\Sigma_t$
in the four-geometry is encoded in the extrinsic curvature: \bea
K_{\mu\nu}=-\frac{1}{2}{\cal L}_n \gamma_{\mu\nu}
    = - {\gamma^\rho}_\mu \nabla_\rho n_\nu \ , \label{kij}
\eea which can also be expressed as \bea
K_{ij}=-\frac{1}{2\alpha}\left( \partial_t \gamma_{ij} -D_i\beta_j
- D_j\beta_i\right) \, , \label{kinematical} \eea where $D_i$ is
the connection associated with $\gamma_{ij}$.

Under the 3+1 decomposition, Einstein equations split in two sets:
evolution and constraint equations. In vacuo, the case that we are
interested in, the evolution equations are
\begin{eqnarray}
\partial_t K_{ij} - {\cal L}_{\beta} K_{ij}
    &=& \alpha \left( R_{ij} - 2 K_{ik} K^k_{\ j}
    + K K_{ij} \right)
    \nonumber\\  &-& D_i D_j \alpha \label{evoleq}
    \ ,
\end{eqnarray} where $K$ is the trace of $K_{ij}$
($K=\gamma^{ij}K_{ij}$). On the other hand, the constraint
equations (respectively, Hamiltonian and momentum constraints) are
expressed as \bea R + K^2 - K_{ij} K^{ij} &=& 0\ ,
\label{Hamconstraint} \\ D_j \left(K^{ij} -\gamma^{ij} K\right)
&=& 0 \label{Momconstraint} \ . \eea

In brief, the Initial Data Problem consists in providing pairs
$(\gamma_{ij},K^{ij})$ that satisfy the constraints
(\ref{Hamconstraint}) and (\ref{Momconstraint}) on an initial
Cauchy surface $\Sigma_t$.

The discussion of the notion of quasi-equilibrium demands a
certain control on the time evolution of the relevant fields. The
Conformal Thin Sandwich (CTS) introduced in Refs.
\cite{Yo98,PfeifY03} is particularly well suited, since it
provides an approach to the Initial Data Problem that consistently
incorporates (a part of) the time derivative of the metric,
together with the lapse and the shift.

The CTS approach starts by conformally decomposing the metric and
the extrinsic curvature, the latter expressed in terms of its
trace $K$ and a traceless part $A^{ij}$, \bea \gamma_{ij} &=&
\Psi^4 \tilde{\gamma}_{ij}\ , \nonumber \label{confdescomp}\\
K^{ij} &=& \Psi^{-4} A^{ij} +\frac{1}{3} K \gamma^{ij} \ . \eea In
this expression the conformal factor $\Psi$ is given by \bea \Psi
\equiv \left(\frac{\gamma}{f}\right)^{\frac{1}{12}} \ , \eea where
$\gamma$ is the determinant of $\gamma_{ij}$ and $f$ is the
determinant of $f_{ij}$, an auxiliary time-independent metric,
$\partial_t f_{ij}=0$, which captures the asymptotics of
$\gamma_{ij}$ \cite{BGGN03}.

Substituting the above decomposition of $\gamma_{ij}$ in relation
(\ref{kinematical}) and taking the trace, we find:\footnote{In the
following expressions, objects with an over-tilde are associated
with the conformal metric $\tilde{\gamma}_{ij}$. They are consistent
with the conformal rescalings in Ref. \cite{BGGN03}, rather than 
with those originally introduced in Refs. \cite{Yo98,PfeifY03}.}
\bea
\partial_t\Psi&=& \beta^i\tilde{D}_i\Psi + \frac{\Psi}{6}
(\tilde{D}_i\beta^i- \alpha K) \label{psidot} \ . \eea This
expression will play an important role when setting the
appropriate boundary conditions in Sec. IV.

With Eq. (\ref{confdescomp}), the Hamiltonian constraint
(\ref{Hamconstraint}) can be written as an elliptic equation for
the conformal factor: \bea \tilde{D}_i\tilde{D}^i \Psi
   = {\Psi\over 8} \tilde R - \Psi^5 \left(
     {1\over 8} \tilde{A}_{ij}A^{ij}
    - {K^2\over 12} \right) \label{psieq} \, ,
\eea whereas the momentum equation is expressed as an elliptic
equation for the shift \bea
  & &{\tilde D}_j {\tilde D}^j \beta^i
    + {1\over 3} {\tilde D}^i {\tilde D}_j \beta^j
    + {\tilde R}^i_{\ \, j} \beta^j
    - ({\tilde L}\beta)^{ij} \tilde{D}_j \ln(\alpha \Psi^{-6})
     \nonumber\\
        & &\hspace{.2cm}=  {4\over 3}\alpha {\tilde D}^i K
         - \tilde{D}_j {\partial_t\tilde{\gamma}}^{ij}
    +   {\partial_t\tilde{\gamma}}^{ij}
    \tilde{D}_j \ln(\alpha \Psi^{-6}) \ ,\label{shifteq}
\eea where $\tilde{A}_{ij}\!\equiv\!
\tilde{\gamma}_{ik}\tilde{\gamma}_{jl}A^{kl}$, $\tilde{D}_i$ is
the connection associated with ${\tilde\gamma}_{ij}$, and
$({\tilde L}\beta)^{ij}\!\equiv\! \tilde{D}^i \beta^j +
\tilde{D}^j \beta^i- \frac{2}{3} \tilde{D}_k \beta^k
\tilde{\gamma}^{ij}$.

Only the conformal part $\tilde{\gamma}_{ij}$ of the metric
$\gamma_{ij}$ encodes dynamical degrees of freedom. This suggests
solving the trace of the evolution equations (\ref{evoleq})
together with the constraints, defining in this way an {\it
enlarged} problem on the initial surface \cite{PfeifY03}. This
additional equation turns out to be elliptic in its dependence on
the lapse $\alpha$, \begin{eqnarray} \!
    & &\hspace*{-.3cm}\tilde{D}_i\tilde{D}^i \alpha + 2
    \tilde{D}_i\ln\Psi \,
    \tilde{D}^i \alpha \nonumber\\
    & &=\,
    \Psi^4 \left\{
    \alpha\left( \tilde{A}_{ij} A^{ij}
    + {K^2\over 3} \right)\!
    + \beta^i \tilde{D}_i K - \partial_tK \right\}\!
    \label{lapseeq} .
\end{eqnarray}
An extra justification to add this equation is because it
straightforwardly permits to impose the condition $\partial_t
K=0$, a good ansatz for quasi-equilibrium. In this extended
problem the constrained parameters are given by $(\Psi, \beta^i,
\alpha)$ and the free data on the initial Cauchy surface are
$(\tilde{\gamma}_{ij},\partial_t\tilde{\gamma}^{ij},K,
\partial_tK)$,
subject to the constraints ${\rm det}(\tilde{\gamma}_{ij}) = f$
and $\tilde{\gamma}_{ij}\partial_t\tilde{\gamma}^{ij}=0$ [in the
strict Initial Data Problem $\frac{\alpha}{\Psi^6}$ is a free
parameter on the initial slice, but here it is constrained owing
to Eq. (\ref{lapseeq})]. Hence, the inner boundary problem
presented in the Introduction reduces to the search for
appropriate boundary conditions for $\Psi$, $\beta^i$, and
$\alpha$ imposed on the horizon.

\section{Isolated Horizons formalism}

In this section we will motivate the introduction of the notion of
Isolated Horizon (IH) and summarize the concepts and definitions
that will be employed in Sec. IV. We will try to provide a
presentation as accessible as possible for a broad community. For
a detailed and more rigorous discussion of the IH formalism see,
e.g., Ref. \cite{ABL01a} or Ref. \cite{AK04}.

The physical scenario that the IH construction attempts to
describe is that of a dynamical space-time containing a black hole
in equilibrium, in the sense that neither matter nor radiation
cross its horizon. This scenario applies as an approximation for
each of the two black holes in a binary before their merger,
provided that they are sufficiently separated, therefore
justifying the relevance of the IH formalism for the Initial Data
Problem of binary black holes.

A very important feature of the IH formalism is its (quasi-)local
character. In our context, the need of a (quasi-)local description
is motivated first by the way in which numerical simulations are
designed from a 3+1 approach, in which we do not have {\it a
priori} control on global space-time properties, and secondly by
the desire of characterizing physical parameters of the black hole
as well as the concept of equilibrium in a (quasi-)local manner.
The notion of apparent horizon, with a local characterization as
an outermost marginal trapped surface \footnote{That is, a surface
${\cal S}$ in $\Sigma_t$ on which the expansions $\theta_{(l)}$
and $\theta_{(k)}$ of the outgoing and ingoing null vectors,
$l^\mu$ and $k^\mu$ respectively, satisfy $\theta_{(l)}=0$ and
$\theta_{(k)}<0$.} in a three-slice, seems an adequate starting
point. However, to include the concept of equilibrium we must
somehow consider the evolution of this two-dimensional surface. In
the (quasi-)equilibrium regime, the notion of the world-tube of an
apparent horizon does in fact make sense (there are no jumps).
Actually, an IH implements the idea that an apparent horizon
associated with a black hole {\it in equilibrium} evolves smoothly
into apparent horizons of the same area, in such a way that the
generated world-tube is a null hypersurface. This null character
encodes the key quasi-equilibrium ingredient, and is essentially
linked to the idea of keeping constant the area of the apparent
horizon.

Inspired by these considerations, the definition of IH tries to
seize the fundamental ingredients of the null world-tube of a
non-expanding apparent horizon. In doing this, the world-tube is
endowed with some additional geometrical structures that are
intrinsic to the null hypersurface \cite{ABL01b}. The specific
amount and nature of these extra structures depend on the physical
problem that one wants to address. This introduces a hierarchy of
structures in the formalism which turns out to be very useful for
keeping track of the hypotheses that are assumed to hold, as will
become evident in Sec. IV.

Before describing these structures, let us emphasize the change of
strategy with respect to Sec. II: while the relevant geometry was
there that of the spacelike three-surface where the initial data
are given, the relevance corresponds now to a null three-geometry.
The combined use of these two complementary perspectives, each of
them suggesting their own natural geometrical objects, will prove
to be specially fruitful.

\subsection{Non-Expanding Horizons}

\subsubsection{Definition}

A first level in the hierarchy of structures entering the IH
formalism is the notion of Non-Expanding Horizon (NEH), which
incorporates the idea of quasi-equilibrium sketched above. We say
that a hypersurface $\Delta$ in a vacuum space-time $({\cal M},
g_{\mu\nu})$ is a NEH if \cite{ABL01b}:
\begin{itemize}
\item [{\it i)}] It is a
null hypersurface with $S^2\times ({\cal I} \subset {\mathbb R})$
topology. That is, there exists a null vector field $l^\mu$ on
$\Delta$, defined up to rescaling, such that $g_{\mu\nu} l^\mu
v^\nu=0$ for all vectors $v^\mu$ tangent to $\Delta$. The
degenerate metric induced on $\Delta$ by $g_{\mu\nu}$ will be
denoted by $q_{\mu\nu}$.

\item [{\it ii)}] The expansion
$\theta_{(l)}\equiv q^{\mu\nu}\nabla_\mu l_\nu$ of any null normal
$l^\mu$ vanishes on $\Delta$.\footnote{Here $q^{\mu\nu}$ is any
tensor satisfying
$q^{\rho\sigma}q_{\rho\mu}q_{\sigma\nu}=q_{\mu\nu}$.}

\item [{\it iii)}] Einstein equations are satisfied on $\Delta$.
\end{itemize}
Matter can be included without problems in the scheme, but we will
focus here on the vacuum case.

\subsubsection{Main consequences for our problem}

Let us first note that the cross-sections ${\cal S}\simeq S^2$ of
the NEH are not necessarily strict apparent horizons since they
are not imposed to be outermost surfaces and no condition is
enforced on the expansion $\theta_{(k)}$ of the ingoing null
vector $k^\mu$ (see footnote 2). Abusing of the language, we will
however refer throughout to the cross-sections as apparent
horizons, a practice ultimately justified in our problem by a
sensible choice of freely specifiable data on the initial surface.
\vspace{.15cm}

{\it a) Constant area.} Owing to the null character of $\Delta$,
any null generator $l^\mu$ defines a {\it natural} evolution on
the hypersurface, in such a way that the area of the apparent
horizons ($a=\int_{\cal S} d^2 V = \int_{\cal S} \sqrt{q} d^2q$,
where we use the natural metric $q_{ab}$ on ${\cal S}$ induced by
$q_{\mu\nu}$) does not change, since ${\cal L}_l \left(\ln
\sqrt{q}\right)=\theta_{(l)}=0$. Therefore, there is a
well-defined notion of radius of the horizon, $R_\Delta \equiv
\sqrt{a}/(4\pi)$.\vspace{.15cm}

{\it b) Surface gravity.} Since $l^\mu$ is null and normal to
$\Delta$, it can be shown to be pre-geodesic and twist free.
Hence,\footnote{The symbol $\stackrel{\Delta}{=}$ denotes equality
on the horizon $\Delta$.} \bea \nabla_l l^\mu \stackrel{
\Delta}{=} \kappa_{(l)} \, l^\mu \ , \label{pregeodesic} \eea
where $\kappa_{(l)}$ is a function on $\Delta$ that will be
referred to as {\it surface gravity} (see Appendix A).
\vspace{.15cm}

{\it c) Second fundamental form $\Theta_{\mu\nu}$ on $\Delta$ and
evolution Killing vector on $\Delta$.} We introduce the second
fundamental form of $\Delta$ \cite{Damou79} \bea
\Theta_{\mu\nu}&\equiv&\frac{1}{2}{P^\alpha}_\mu {P^\beta}_\nu
{\cal L}_l q_{\alpha\beta}= \frac{1}{2}{q^\alpha}_\mu
{q^\beta}_\nu {\cal L}_l q_{\alpha\beta} \ , \label{sec} \eea
where ${P^\alpha}_\beta=\delta^\alpha_\beta +k^\alpha l_\beta$ and
${q^\alpha}_\beta=\delta^\alpha_\beta +k^\alpha l_\beta + l^\alpha
k_\beta$, with $k_\mu l^\mu=-1$. This is an essentially
two-dimensional object {\it living} on apparent horizons, such
that $\theta_{(l)}={\Theta^\mu}_\mu$ while \bea \Theta_{ab}
&\equiv& \frac{1}{2} \theta_{(l)} q_{ab} + {\sigma_{(l)}}_{ab}
\label{thetasigma} \eea defines the shear ${\sigma_{(l)}}_{ab}$
associated with $l^\mu$. Since Einstein equations hold on
$\Delta$, so does Raychaudhuri equation. In vacuo, and since the
twist of $l^\mu$ cancels, it has the form \bea {\cal
L}_l\theta_{(l)}=\kappa_{(l)}\theta_{(l)}
-\frac{1}{2}\theta_{(l)}^2 -{\sigma_{(l)}}_{ab}{\sigma_{(l)}}^{ab}
\ . \label{Raychaudhuri} \eea The vanishing of $\theta_{(l)}$
throughout $\Delta$ (so that, in particular, ${\cal
L}_l\theta_{(l)}\stackrel{\Delta}{=}0$) implies then the vanishing
of the shear ${\sigma_{(l)}}_{ab}$. As a consequence,
$\Theta_{\mu\nu}\stackrel{\Delta}{=} 0 $. From Eq. (\ref{sec}) we
then see that, on $\Delta$, $q_{\mu\nu}$ is Lie-dragged by the
null vector $l^\mu$. Therefore, although in general there is no
Killing vector of the full space-time, the induced metric on
$\Delta$ admits an intrinsic Killing symmetry. This fact extracts
from the stronger notion of Killing horizon \cite{Carte69} the
relevant part for our problem. \vspace*{.15cm}

{\it d) Connection $\omega_\mu$.} The vanishing of
$\Theta_{\mu\nu}$ and the fact that $l^\mu$ is normal to $\Delta$
suffice to define a one-form $\omega_\mu$ intrinsic to $\Delta$,
such that \bea v^\nu\nabla_\nu l^\mu \stackrel{\Delta}{=}
v^\nu\omega_\nu \,l^\mu \label{omega} \eea for any vector $v^\mu$
tangent to $\Delta$. This one-form provides a strategy for
computing $\kappa_{(l)}$ in Eq. (\ref{pregeodesic}): \bea
\kappa_{(l)}\,\stackrel{\Delta}{=}\, l^\mu\omega_\mu \label{kappa}
\ . \eea In addition, we will see that it plays a central role in
introducing the next level of the IH hierarchy of structures.

{\it e) Transformations under rescaling of $l^\mu$.} For later
applications, let us also summarize the transformation of the main
geometrical objects under a rescaling of $l^\mu$ by a function
$\lambda$ on $\Delta$. Under a change $l^\mu\to \lambda \; l^\mu$
we find \begin{eqnarray} q_{\mu\nu} &\to& q_{\mu\nu}\ , \quad
\quad \quad \ \omega_\mu \to \omega_\mu + {P^\nu}_{\mu} \nabla_\nu
\ln \; \lambda \ , \nonumber\\ \Theta_{\mu\nu} &\to& \lambda
\;\Theta_{\mu\nu} \ , \quad \quad \kappa_{(l)}\to \lambda \;
\kappa_{(l)}+ l^\mu\nabla_\mu \lambda \label{rescalings} \ .
\end{eqnarray}
It is obvious from these expressions that the characterization of
NEH does not depend on the rescaling of $l^\mu$.

\subsubsection{ 3+1 perspective of Non-Expanding Horizons}

As discussed above, we want to cope with intrinsic evolution
properties of apparent horizons. However, in contrast with the
previous discussion on NEH, in our Initial Data Problem we only
dwell on a given spatial slice (at most, on two infinitesimally
closed slices in the CTS), not on the whole world-tube. Therefore,
we must find a procedure to characterize an apparent horizon as a
section of an IH by only using information on the initial spatial
slice. From the NEH definition, a NEH of infinitesimal width is
implemented if, together with the condition $\theta_{(l)}=0$, we
are able to enforce ${\cal L}_l \theta_{(l)}=0$ on the initial
sphere ${\cal S}$. The Raychaudhuri equation (\ref{Raychaudhuri})
leads then to the characterization given in Ref. \cite{DKSS02},
that can be expressed as:

{\it The infinitesimal world-tube of an apparent horizon ${\cal
S}$ is a NEH if and only if the shear ${\sigma_{(l)}}_{ab}$ of the
outgoing null vector vanishes on ${\cal S}$}.

Of course, if we want to extend the NEH character to a finite
world-tube, we need to find a way to impose these conditions on a
finite evolution interval, something that is not possible in the
Initial Data Problem. At least, this {\it instantaneous} notion of
equilibrium must be completed with a proper choice of dynamical
content in the free data on the initial Cauchy surface.
Summarizing, we see from Eq. (\ref{thetasigma}) that the condition
that we must impose on the sphere ${\cal S}$ in $\Sigma_t$ in
order to have a section of a NEH is: \bea \Theta_{ab}\, |_{_{\cal
S}}=\; 0 \, \label{Theta0} \eea where the symbol $|_{_{\cal S}}$
stands for evaluation on ${\cal S}$.

\subsection{Weakly Isolated Horizons}

A NEH describes a minimal notion of quasi-equili\-brium, but it is
not rich enough for allowing the assignment of well-defined
physical parameters to the black hole. In order to do so, we must
endow the horizon with extra structure. Noting that the key
property of the NEH is that $l^\mu$ is a Killing vector of the
metric induced on the horizon, a way to introduce new structure
con\-sists in enforcing that other objects are Lie-dragged by
$l^\mu$.

A simple choice in this sense, that permits a Hamiltonian analysis
leading to (quasi-)local physical quantities associated with the
black hole, is to demand that ${\cal L}_l
\omega_\mu\,\stackrel{\Delta}{=}\,0$. However, the transformation
rule of $\omega_\mu$ in Eq. (\ref{rescalings}) precludes this
condition to hold for every null normal $l^\mu$. Nonetheless, a
consistent way to impose it is by introducing the notion of Weakly
Isolated Horizon: \footnote{The one-form $\omega_\mu$ encodes some
of the components of a connection $\hat{\nabla}$ on $\Delta$
compatible with the degenerate metric $q_{\mu\nu}$ \cite{Damou79}.
This connection $\hat{\nabla}$ is in fact unique as a consequence
of the NEH definition. The stronger condition $[{\cal
L}_l,\hat{\nabla}] = 0$ defines a Strongly Isolated Horizon, a
much more rigid structure.}

{\it A Weakly Isolated Horizon (WIH) is a NEH endowed with an
equivalence class $[l^\mu]$ of null normals ($l'^\mu$ and $l^\mu$
belong to the same class if and only if $l'^\mu = c \;l^\mu$ with
$c$ a positive constant) such that:} \bea {\cal L}_l\omega_\mu
\stackrel{\Delta}{=} 0 \ . \label{Lw} \eea This condition turns
out to be equivalent to (see Appendix A)\bea d
\left(\kappa_{(l)}\right)\stackrel{\Delta}{=} 0 \ ,
\label{WIHcharac} \eea so that the zeroth law of black hole
thermodynamics, $\kappa_{(l)}= {\rm constant}$, characterizes the
WIH notion.

It is worth commenting that, given a NEH, it is always possible to
select a class of null normals $[l^\mu]$ such that $\Delta$
becomes a WIH. Actually there exists an infinite freedom in the
construction of the WIH structure \cite{ABL01b}. Namely, if the
surface gravity $\kappa_{(l)}$ is a (non-vanishing) constant for a
certain class of null normals $[l^\mu]$, the same happens for any
of the classes obtained by the non-constant rescaling \bea
\hat{l}^\mu\stackrel{\Delta}{=}[1+B(\theta,\phi)
e^{-\kappa_{(l)}v}]\,l^\mu\, ,\label{scalewih}\eea where
$B(\theta,\phi)$ is an arbitrary function on ${\cal S}$ and $v$ is
a coordinate on $\Delta$ compatible with $l^\mu$, i.e., ${\cal
L}_lv\stackrel{\Delta}{=}1$. In fact, the above rescaling does not
modify the constant value of the surface gravity [this follows
from the transformation rule for $\kappa_{(l)}$ in Eq.
(\ref{rescalings})].

Since it is always possible to find WIH structures on a given NEH,
the WIH concept does not correspond to a real restriction on the
physics of the system. However it does impose a restriction on the
spacetime slicing by the hypersurfaces $\Sigma_t$ introduced in
Sec.~\ref{s:CFS} if we tie $l^\mu$ to $t$, i.e. if we impose that
there is a member $l^\mu$ of the WIH class $[l^\mu]$ such that
${\cal L}_l t \stackrel{\Delta}{=}1$. We call such a slicing a
{\em WIH-compatible slicing}.

The derivation of the mass and angular momentum expressions for a
WIH using Hamiltonian techniques is beyond the scope of this work
(see Refs. \cite{AFK00,ABL01a}). Here, we will simply extract
those points which are relevant for our analysis. The general idea
is to characterize physical parameters as conserved quantities of
certain transformations that are associated with symmetries of the
WIH. A vector field $V^\mu$ tangent to $\Delta$ is said to be a
symmetry of the particular WIH under consideration if it preserves
its equivalence class of null normals, the metric $q_{\mu\nu}$,
and the one-form $\omega_{\mu}$, namely \bea \!\!{\cal L}_{V}
l^\mu \stackrel{\Delta}{=}{\rm constant} \cdot l^\mu , \ \ \ {\cal
L}_{V} q_{\mu\nu} \stackrel{\Delta}{=} 0 \ , \ \ \ {\cal L}_{V}
\omega_\mu \stackrel{\Delta}{=} 0 \, . \label{WIHsym} \eea In Sec.
IV we will be interested in non-extremal black holes, for which
$\kappa_{(l)}\neq 0$. In that case, the general form of a WIH
symmetry is \cite{ABL01a} \bea V^\mu = c_{V} l^\mu + b_{V} S^\mu
\, ,\label{WIHsymmetry} \eea where $c_{V}$ and $b_{V}$ are
constant on $\Delta$ and $S^\mu$ is an isometry of the apparent
horizon ${\cal S}$.

The definition of the conserved quantities goes first through the
construction of an appropriate phase space for the problem and
then through the analysis of canonical transformations on this
phase space \cite{ABL01a}. An important point is that the relevant
transformations are generated by diffeomorphisms in space-time
whose restriction to the horizon $\Delta$ are symmetries of the
WIH in the sense of (\ref{WIHsym}).

\subsubsection{Angular momentum}

In order to define a conserved quantity that we can associate with
a (quasi-)local angular momentum, we assume that there exists an
azimuthal symmetry on the horizon $\Delta$ (actually, this
hypothesis can be relaxed; see in this sense Ref. \cite{AK04}).
Therefore, we assume the existence of a vector $\varphi^\mu$
tangent to ${\cal S}\subset\Delta$, which is a $SO(2)$ isometry of
the induced metric $q_{ab}$ with $2 \pi$ affine length.

The conserved quantity associated with an extension of
$\varphi^\mu$ to the space-time is given by \cite{ABL01a} \bea
\!\!J_\Delta = -\frac{1}{8\pi G}\int_S \varphi^\mu \omega_\mu \;
d^2 V = \frac{1}{8\pi G}\int_{\cal S}\! s^i\varphi^j K_{ij} \; d^2
V \, , \label{angmomentum} \eea where for convenience we have
expressed it in terms of objects in the 3+1 decomposition. In
particular, $s^i$ is the outward (pointing towards spatial
infinity) unit vector field in $\Sigma_t$ normal to the apparent
horizon ${\cal S}$.

\subsubsection{Mass and boundary condition for $t^\mu$}

The definition of the mass is related to the choice of an
evolution vector $t^\mu$ with appropriate boundary conditions,
namely that $t^\mu \to (\partial_t)^{\mu}$ at spatial infinity and
$t^\mu \to l^\mu - \Omega_t \varphi^\mu$ with $l^\mu \in[l^\mu]$
and $\Omega_t$ constant on the horizon (note that
$t^\mu\!\!\mid_\Delta$ is a WIH symmetry). The determination of
the mass expression proceeds in two steps.

First, the vector $t^\mu$ has to satisfy certain conditions to
induce a canonical transformation on the phase space. This turns
out to be equivalent to the first law of black hole thermodynamics
\cite{AFK00,ABL01a}, whose practical consequence for us is that
the mass $M$, the surface gravity $\kappa_{(l)}$, and the angular
velocity $\Omega_t$ depend only on the radius $R_\Delta$ and the
angular momentum $J_\Delta$ of the black hole,
\begin{eqnarray} \;M&=&M(R_\Delta, J_\Delta) \ ,\nonumber\\
\kappa_{(l)}&=&\kappa_{(l)}(R_\Delta, J_\Delta) \ ,\nonumber\\
\Omega_t &=& \Omega_t (R_\Delta, J_\Delta) \ , \end{eqnarray} but
without determining the specific functional form. It is worth
emphasizing that this dependence on $R_\Delta$ and $J_\Delta$
(though arbitrary in principle) must be the same for all solutions
to the Einstein equations containing a WIH.

In a second step, this dependence is fixed to coincide with that
found in the stationary Kerr family of black holes. This is not an
arbitrary choice, but a normalization consistent with the
stationary solutions. Technically, this is accomplished by
requiring that, at the horizon $\Delta$, $t^\mu +
\Omega_t\varphi^\mu$ reproduces just the null normal (in the
considered class $[l^\mu]$) whose surface gravity equals that of
the Kerr case, something that is always possible in the
non-extremal situation via a constant rescaling. This singles out
a vector $t_o^\mu$, satisfying \bea t_o^\mu +
\Omega_{_{Kerr}}(R_{\Delta},J_{\Delta}) \varphi^\mu
\stackrel{\Delta}{=} c\; l^\mu \equiv l_o^\mu \label{to} \eea with
\bea c \equiv
\frac{\kappa_{_{Kerr}}(R_{\Delta},J_{\Delta})}{\kappa_{(l)}}
\label{c} \ , \eea as the evolution vector used for the derivation
of the mass formula. The final expressions obtained in this way
for the physical parameters of the horizon are: \bea M_\Delta
&\equiv&M_{_{Kerr}}(R_{\Delta},J_{\Delta}) =
\frac{\sqrt{R_{\Delta}^4 + 4G^2 J_{\Delta}^2}}{2 G R_{\Delta}}\
,\nonumber
\\ \kappa_\Delta&\equiv&\kappa_{_{Kerr}}(R_{\Delta},J_{\Delta})
=\frac{R_{\Delta}^4 - 4 G^2 J_{\Delta}^2} {2 R_{\Delta}^3
\sqrt{R_{\Delta}^4 + 4G J_{\Delta}^2}}\ , \nonumber
\\ \Omega_\Delta&\equiv&\Omega_{_{Kerr}}(R_{\Delta},J_{\Delta})
=\frac{2 G J_{\Delta}}{R_{\Delta} \sqrt{R_{\Delta}^4 + 4G
J_{\Delta}^2}} \label{physparam}\ . \eea

\section{Derivation of the boundary conditions}

We are now in an adequate situation to derive boundary conditions
for the elliptic equations in Sec. II. In doing so, we adopt a
coordinate system $(t,x^i)$ stationary with respect to the
horizon, in the sense that the null tube $\Delta$ can be
identified as the hypersurface $r(x^i)={\rm constant}$ for a
certain function $r$ which is independent of $t$. It can be shown
that this happens if and only if $t^\mu$ is chosen tangent to
$\Delta$, i.e. $l_\mu t^\mu\stackrel{\Delta}{=}0$.

Since we want to have a notion of angular momentum for the black
hole, following the discussion in Subsec. III.B.1 we make the
hypothesis that {\it our physical regime permits the imposition of
an axial isometry $\varphi^a$ on ${\cal S}\simeq S^2
\subset\Delta$}. Even though this is a strong physical hypothesis
(especially when having in mind binary black holes), we must
emphasize that the bulk space-time will still be generally
dynamical in an arbitrarily close neighborhood of the horizon and
that $\varphi^{\mu}$ does not need to extend to an isometry there.

To construct the equilibrium black hole on ${\cal S}$, we follow
the steps dictated by the hierarchy of the IH formalism.

\subsection{Adapting the evolution vector to the horizon}

Aiming at imposing the NEH structure, but already motivated by the
boundary condition for the evolution vector selected by the
determination of $t_o^\mu$, we adapt $t^\mu$ to the horizon by
relaxing to a NEH the particular WIH structure implicit in Eq.
(\ref{to}). That is, we only impose \bea t^\mu + \Omega_\Delta \;
\varphi^\mu \stackrel{\Delta}{\sim} {l}^\mu \, ,\eea where the
proportionality needs not be given by a constant on $\Delta$.
Using the proportionality \bea l^\mu \sim \left(n^\mu +
s^\mu\right) \eea (where $s^\mu$ is again the outward unit spatial
vector normal to ${\cal S}$) and the decomposition
(\ref{alphabeta}) of $t^\mu$ in terms of the lapse and shift, we
conclude \bea \beta^i \stackrel{\Delta}{=} \alpha s^i -
\Omega_\Delta \; \varphi^i \ , \label{bcshift} \eea from which
boundary conditions for the shift on ${\cal S}$ immediately
follow.

Actually the choice of stationary coordinates with respect to the
horizon automatically leads to the expression $\beta^i
\stackrel{\Delta}{=} \alpha s^i - W^i$, where $W^i$ is an
arbitrary vector tangent to ${\cal S}$. So, here we enforce $W^i$
to be {\it precisely} $\Omega_\Delta \,\varphi^i$, a choice that
will in fact simplify the imposition of the NEH
structure.\footnote{Although the derivation of $J_\Delta$ actually
involves a WIH structure, its expression can be shown to be
already well-defined for a NEH.}

\subsection{Non-Expanding Horizon condition}

We now properly impose the NEH condition. As mentioned in Subsec.
III.A.3, in this Initial Data Problem we demand ${\cal S}$ to be a
slice of a NEH of infinitesimal width. For this, we impose
condition (\ref{Theta0}). Owing to the rescaling property
(\ref{rescalings}) of $\Theta_{\mu\nu}$ under an arbitrary (not
necessarily constant) rescaling of $l^\mu$, and taking advantage
of the $t^\mu$ adaptation to the horizon implemented by the shift
boundary conditions, we can write \bea {q^\alpha}_\mu
{q^\beta}_\nu {\cal L}_{_{[t+\Omega_\Delta
\varphi]}}q_{\alpha\beta}\,\stackrel{\Delta}{=}\,0 \ . \label{Ltq}
\eea In our stationary coordinates with respect to the horizon,
this simply reads \bea
0\,\stackrel{\Delta}{=}\,2\Theta_{ab}\,\stackrel{\Delta}{=}\,
\partial_t q_{ab} + \Omega_\Delta\; {\cal L}_\varphi q_{ab} \ .
\label{condgen} \eea But, under our hypothesis about the existence
of an axial isometry on the horizon, the second term must vanish
on its own: ${\cal L}_\varphi q_{ab}\,\stackrel{\Delta}{=}\,0$.
Using $q_{ab}\stackrel{\Delta}{=}\gamma_{ab}$ (for angular
co\-variant components) we find \bea
\partial_t\gamma_{ab}\stackrel{\Delta}{=} 0 \ ,
\ \ \ \ \ \ \ \ \ {\cal L}_\varphi\gamma_{ab} \stackrel{\Delta}{=}
0 \ . \label{condgenpart} \eea In particular, the restrictions
must hold on ${\cal S}$. These are the NEH boundary conditions.
Note that their simple form depends critically on the specific
choice made for $W^i$ in the previous subsection.

Using now the conformal decomposition of the metric, these
conditions translate into \bea
\left(4\tilde{\gamma}_{ab}\partial_t{\Psi}+\Psi\partial_t
\tilde{\gamma}_{ab}\right)|_{_{\cal S}} \!=0\, ,&& \
\label{NEHtotal1}
\\ \left(4\tilde{\gamma}_{ab}{\cal L}_\varphi\Psi+
\Psi{\cal L}_\varphi\tilde{\gamma}_{ab}\right)|_{_{\cal S}} \!=0\,
.&& \ \label{NEHtotal2} \eea The crucial feature, and the ultimate
reason for using the CTS, is that these conditions can be
satisfied by an appropriate choice of the free data
$\tilde{\gamma}_{ab}$ and $\partial_t{\tilde{\gamma}}_{ab}$.
Condition (\ref{NEHtotal2}), expressing the axial symmetry of the
horizon, must be enforced by a self-consistent selection of the
free data $\tilde{\gamma}_{ab}$ on ${\cal S}$ ($\Psi$ is a
functional of $\tilde{\gamma}_{ij}$). Regarding (\ref{NEHtotal1}),
if we first take its trace with respect to the {\it conformal}
counterpart of the metric $q_{ab}$ induced on ${\cal S}$,
$\tilde{q}_{ab}\equiv \Psi^{-4} q_{ab}$ (satisfying
$\tilde{q}^{ac}\tilde{\gamma}_{cb}=\delta^a_b$) and then use Eq.
(\ref{psidot}) for $\partial_t \Psi$, we find (calling $tr_{_{\cal
S}}\dot{\tilde{\gamma}}\equiv
\tilde{q}^{ab}\partial_t\tilde{q}_{ab}\,\stackrel{\Delta}{=}\,
\tilde{q}^{ab}\partial_t\tilde{\gamma}_{ab}$) \bea
\left.\left[\beta^i\tilde{D}_i\Psi + \frac{\Psi}{6} \left( {\tilde
D}_i \beta^i - \alpha K + \frac{3}{4}tr_{_{\cal
S}}\dot{\tilde{\gamma}} \right)\right]\right|_{_{\cal S}}\!\!= 0 \
. \label{bcpsi} \eea In addition, from Eq. (\ref{NEHtotal1}) it
follows that the $\tilde{q}$-traceless part of
$\partial_t{\tilde{\gamma}}_{ab}$ must vanish. Therefore, on the
boundary ${\cal S}$, this part of the free data has the form \bea
\left.\left(\partial_t\tilde{\gamma}_{ab}- \frac{1}{2} tr_{_{\cal
S}}\dot{\tilde{\gamma}}
\;\tilde{\gamma}_{ab}\right)\right|_{_{\cal S}}\!\!=0\, .
\label{trazatheta}\eea

Condition (\ref{bcpsi}) is an inner boundary condition for $\Psi$.
Since we have imposed on ${\cal S}$ the Dirichlet boundary
conditions (\ref{bcshift}) on $\beta^i$, we have no direct control
on the sign of $\tilde{D}_i\beta^i$ there. In order to guarantee
the positivity of $\Psi$ via the application of a maximum
principle, the factor multiplying $\Psi$ in Eq. (\ref{bcpsi})
should be non-negative. The analytical study of this issue goes
beyond the present geometrical derivation (see Ref. \cite{Da04}
for a discussion on this point in a related context). We simply
comment that the choice of free data for $tr_{_{\cal
S}}\dot{\tilde{\gamma}}$ [and, more indirectly, that of the radial
components of the free data $\tilde{\gamma}^{ij}$, which determine
$s^i=\gamma^{ri}/\sqrt{\gamma^{rr}}$ in condition (\ref{bcshift})]
could play a key role in ensuring that Eq. (\ref{bcpsi}) is a
well-posed condition.

Except for the implicit use of a well-defined concept of angular
momentum based on the WIH formalism, the notion of
quasi-equilibrium provided by the NEH structure has proved to be
sufficient to set boundary conditions for the Initial Data
Problem, since it prescribes boundary values for $\Psi$
(Hamiltonian constraint) and $\beta^{i}$ (momentum constraint). If
this is the problem that we want to solve, we can stop here.
However, if we want to solve also the trace of the evolution
equations, we need to find appropriate boundary conditions for the
lapse. We will show how the existence of a WIH structure can be
exploited with that aim.

\subsection{Weakly Isolated Horizon condition}

As we have commented, given a NEH one can always find a class of
null normals so that it becomes weakly isolated. In fact, the
determination of this class is not unique, but there exists an
infinite freedom of choice. In this subsection, we will first
discuss the restrictions on the lapse function that follow from
the introduction of a WIH-compatible slicing and then employ the
freedom in the choice of WIH to suggest possible boundary
conditions for the lapse that are specially suitable for numerical
integration.

Let us start by choosing $l^\mu_o$ as the representative of the
class of null normals for the WIH, $[l^\mu]$. The inner boundary
condition (\ref{to}) employed in the determination of the mass
formula then singles out an evolution vector $t_o^\mu$ on the
horizon. We proceed as in Subsec. IV.A, but imposing $t^\mu$ to
coincide exactly with $t_o^\mu$, therefore demanding that the
$\left( \Sigma_t \right)$ foliation constitutes a WIH-compatible
slicing. According to the characterization (\ref{WIHcharac}) of
the WIH notion, the surface gravity $\kappa_{(l)}$ is constant. We
further assume $\kappa_{(l)}\neq 0$, thus restricting the analysis
to the non-extremal case. We write $l^\mu_o$ in terms of $3+1$
objects, \bea l^\mu_o = \alpha \tilde{l}^\mu \equiv \alpha(n^\mu +
s^\mu) \, , \label{scaling} \eea where we have explicitly defined
the vector $\tilde{l}^\mu$. Again, the introduction of the lapse
and shift decomposition for $t^\mu_o$ in Eq. (\ref{to}) leads to
the boundary conditions (\ref{bcshift}) for the shift. In order to
analyze the conditions on $\alpha$, we calculate the expression
for $\kappa_{(l_o)}$. We proceed in several steps:
\begin{enumerate}
\item Contracting Eq. (\ref{omega}), particularized
to the one-form $\tilde{\omega}_\mu$ associated with
$\tilde{l}^\mu$, with the ingoing null covector $\tilde{k}_\mu=
(n_\mu-s_\mu)/2$ and expanding the resulting expression, we find
for any vector $v^\mu$ tangent to $\Delta$, \bea v^\mu
\tilde{\omega}_\mu \stackrel{\Delta}{=} v^\mu s^\nu\nabla_\mu
n_\nu \ . \eea Employing the definition of the extrinsic curvature
(\ref{kij}) and the identity $n^\rho\nabla_\rho
n_\nu={\gamma^\rho}_\nu\nabla_\rho \,\ln \, \alpha$, we get \bea
v^\mu \tilde{\omega}_\mu \stackrel{\Delta}{=} - v^\mu s^\nu
\left(K_{\mu\nu} +n_\mu\gamma^\rho_\nu\nabla_\rho
\,\ln\,\alpha\right) \, \label{vomega}. \eea
\item
Taking $\tilde{l}^\mu$ as the tangent vector $v^{\mu}$ and
remembering expression (\ref{kappa}), we obtain \bea
\kappa_{(\tilde{l})} \,\stackrel{\Delta}{=}\,
\tilde{l}^\mu\tilde{\omega}_\mu \stackrel{\Delta}{=} s^\mu
\nabla_\mu \,\ln\,\alpha -s^\mu s^\nu K_{\mu\nu} \ .
\label{kaltil} \eea
\item Recalling then the transformation (\ref{rescalings}) of
the surface gravity under a rescaling of the null normal, \bea
\kappa_{(l_o)} \stackrel{\Delta}{=} \alpha (s^\mu \nabla_\mu \,
\ln\alpha - s^\mu s^\nu K_{\mu\nu}) + l_o^\mu\nabla_\mu \ln \alpha
\,.
  \label{kace}
\eea
\item Finally, imposing that
$\kappa_{(l_o)}$ equals $\kappa_{_{Kerr}}(R_\Delta,J_\Delta)$, we
find \bea \kappa_{_{Kerr}}(R_\Delta,J_\Delta)\stackrel{
\Delta}{=}s^i D_i \alpha - s^i s^j K_{ij}\, \alpha + {\cal
L}_{l_o} \ln\alpha \, .\label{alphaevol} \eea
\end{enumerate}
This restriction, arising from the WIH-compatible slicing
condition, can be regarded as an evolution equation for the lapse
on the horizon. Properly speaking, it is not a boundary condition
for the initial data, because it contains the derivative of the
lapse in the direction of $l_o^{\mu}$.

Actually, by exploiting the freedom of choice in the WIH
structure, one can freely set the value of the lapse on the
initial section of the horizon, ${\cal S}$. This is a consequence
of the fact that, from Eq. (\ref{scalewih}) and the relation
$l_o^{\mu}=\alpha\tilde{l}^{\mu}$, a change of WIH structure
results in a rescaling of the lapse:
\bea\hat{\alpha}\stackrel{\Delta}{=}\left[1+B(\theta,\varphi)
e^{-\kappa_{_{Kerr}}(R_\Delta, J_\Delta) \, t }\right] \alpha \, .
\label{alphawih}\eea Since ${\cal S}$ can be identified, e.g.,
with the section $t=0$ of $\Delta$, the initial value of the lapse
on the horizon gets multiplied by an arbitrary positive function
on the sphere. Therefore, it can be chosen at convenience, at
least as far as the demand for a WIH-compatible slicing is
concerned. In fact, if we were solving an evolution problem by
following a constrained scheme (see e.g. Ref. \cite{BGGN03}), this
initial choice for $\alpha$ together with Eq. (\ref{alphaevol})
might be used to set inner boundary conditions for the lapse at
each time step.

Once realized the freedom in the choice of the initial value for
$\alpha$ on ${\cal S}$ that follows from the dynamical character
of Eq. (\ref{alphaevol}) on $\Delta$, one may ask whether it is
possible to benefit from this arbitrariness and put forward a
particular proposal for the choice that could be considered
specially advantageous. In this sense, one would like to ensure
that, under evolution on $\Delta$, the lapse will neither increase
exponentially nor decrease to (zero or) negative values.
Apparently, the best way to favor this, at least locally, is to
pick up, among the infinite WIH structures, that in which the Lie
derivative of the lapse with respect to the null normal vanishes
initially: $({\cal L}_{l_o}\alpha)\!\!\mid_{_{\cal S}}=0$. Using
Eq. (\ref{alphawih}), one can prove under very mild assumptions
that such a choice of WIH structure exists. Adopting it, Eq.
(\ref{alphaevol}) becomes a true boundary condition for the lapse
on ${\cal S}$: \bea \left.\left[s^i D_i \alpha - s^i s^j K_{ij}\,
\alpha\right]\right|_{_{\cal
S}}=\kappa_{_{Kerr}}(R_\Delta,J_\Delta)\, .\label{bclapse} \eea

Note in fact that, to deduce this condition, one only needs to
demand [in the passage from Eq. (\ref{kace}) to Eq.
(\ref{alphaevol})] that $\kappa_{(l_o)}$ coincides with the
constant $\kappa_{_{Kerr}}(R_\Delta,J_\Delta)$ on ${\cal S}$, and
not on the whole of $\Delta$, because one finally restricts his
attention just to the initial section of the horizon. As a
consequence, and in contrast with the situation found for the NEH
conditions, the above prescription for the lapse on the boundary
is only a {\it necessary} condition for specifying a WIH of
infinitesimal width. The extra condition that $\kappa_{(l_o)}$ be
constant in the rest of $\Delta$, namely $({\cal
L}_{l_o}\kappa_{(l_o)})|_{_{\cal S}}\!=0$, would involve the
evolution equations and the second time derivative of the lapse,
and therefore cannot be imposed in terms of the initial data.

Finally, we comment that an alternative way of dealing with the
WIH condition $\kappa_{(l)}={\rm constant}$ would consist in
choosing {\it a priori} the values of $\alpha$ and ${\cal
L}_{l_o}\alpha$ on ${\cal S}$ and then interpreting Eq.
(\ref{alphaevol}) as a constraint on the free data on the inner
boundary.

\subsection{Binary quasi-circular orbits}

In the previous subsections we have characterized the
quasi-equilibrium state of each horizon exclusively in local
terms. However, the study of a binary black hole in quasi-circular
orbits requires, in addition, a global notion of
quasi-equilibrium. In the general case, such a global
quasi-stationary situation is described by the existence of a
global quasi-Killing vector $L^\mu$. In the binary black hole
case, this is a helical vector (see Refs. \cite{FUS02,GGB02}) that
Lie drags the horizons, i.e. $L^\mu\!\!\mid_\Delta$ is tangent to
each horizon $\Delta$. Imposing asymptotic flatness, we have at
spatial infinity $L^\mu\to t^\mu_{\infty}+ \Omega_{orb}
\;\phi^\mu_{\infty}$, where $t^\mu_{\infty}$ and
$\phi^\mu_{\infty}$ are vectors associated with an asymptotic
inertial observer and $\Omega_{orb}$ is the orbital angular
velocity. We can adapt the coordinate system, introducing an
evolution parameter $t'$ such that $L^\mu=(\partial_{t'})^\mu$. In
such a case, with the 3+1 decomposition $L^\mu=\alpha' n^\mu+
\beta'^\mu$, outer {\it corotating} boundary conditions follow
\bea \!\lim_{r\to \infty}\!\beta'^i\approx
\Omega_{orb}\,\phi_\infty^i \, ,\ \ \ \lim_{r\to
\infty}\!\alpha'\approx 1 \, ,\ \ \ \lim_{r\to
\infty}\!\Psi\approx 1 \, . \label{outerbc} \eea In these
coordinates, one chooses the {\it time derivative} part of the CTS
free data to vanish \bea
\partial_{t'}\tilde{\gamma}_{ij}=0 \ , \ \ \ \ \partial_{t'}K=0 \
.\label{quasiequil}\eea

In the general case, $L^\mu|_{\Delta}$ and $t_o^\mu$ defined in
Eq. (\ref{to}) do not coincide. Since they are both tangent to the
horizon, \bea L^\mu \stackrel{\Delta}{=} \rho\; t_o^\mu + \chi^\mu
\ \ , \label{chi} \eea where $\rho$ is a scaling factor and
$\chi^\mu$ is tangent to $\Delta$ with $\chi^\mu n_\mu=0$. As a
consequence, if we adapt coordinates to $L^\mu$, hence using $t'$,
the expressions given in the previous subsections must be
corrected. We will discuss two possibilities.

\medskip

{\it a) Corotating coordinate system (fully adapted to $L^\mu$)}.
From Eqs. (\ref{to}) and (\ref{chi}) we can write \bea L^\mu +
\rho\; \Omega_\Delta \varphi^\mu - \chi^\mu \stackrel{\Delta}{=}
\rho\; l_o^\mu \ .\label{Klo} \eea A natural ansatz for $\chi^\mu$
is given by $\Omega_{orb} \;\phi^\mu$, where $\phi^\mu$ is the
azimuthal vector tangent to each horizon and associated with the
normal direction to the orbital plane. Since $\Omega_\Delta$
provides a well-defined notion of rotation angular velocity, we
can define the corotating physical regime in an intrinsic way as
the case with $\rho=1$ and $ \Omega_{orb}=\Omega_\Delta$, from
which $L^\mu\stackrel{\Delta}{=}l_o^\mu$ follows. More generally,
proceeding as in Subsec. IV.A we find \bea \beta'^i
\stackrel{\Delta}{=} \alpha' s^i - \rho\;\Omega_\Delta \;
\varphi^i \ + \chi^i \ . \label{bcshiftorb} \eea

Imposing the axial symmetry on each horizon $\Delta$, we deduce
again condition (\ref{NEHtotal2}). Defining \bea \eta^\mu \equiv
\rho\;\Omega_\Delta\; \varphi^\mu -\chi^\mu \ ,\label{eta} \eea
the requirements $\Theta_{ab}\stackrel{\Delta}{=}0$ and
$\partial_{t'}\tilde{\gamma}_{ij}=0$ [see Eq. (\ref{quasiequil})]
lead then to the conditions: \bea
\hspace*{-.8cm}&&\left.\left[\beta'^i\tilde{D}_i\Psi \!+\!
\frac{\Psi}{6} \left(\tilde{D}_i\beta'^i \!- \alpha' K \right)\!+
\!\frac{1}{8\Psi^3} \tilde{q}^{cd}{\cal
L}_{\eta}\gamma_{cd}\right]\!\right|_{_{\cal S}}\!\!=\!0\,
,\label{bcpsiorb}
\\ \hspace*{-.8cm}&&
\left.\left[\frac{1}{2} \left(\tilde{q}^{cd}{\cal L}_{\eta}
\gamma_{cd}\right) \tilde{\gamma}_{ab}-{\cal L}_{\eta}\gamma_{ab}
\right]\!\right|_{_{\cal S}}\!\!=0 \, , \ \ \label{trazathetaorb}
\eea which replace Eqs. (\ref{bcpsi}) and (\ref{trazatheta}),
respectively. Finally, the condition for the lapse is still
derived as in Subsec. IV.C. From Eq. (\ref{chi}) it follows that
$\alpha'\stackrel{\Delta}{=} \rho\, \alpha$, and hence it can be
shown \bea \left.\left[\frac{s^i D_i\alpha'- s^i s^j K_{ij}\,
\alpha'}{\rho}\right]\right|_{_{\cal S}}\!\!\!=
\kappa_{_{Kerr}}(R_\Delta,J_\Delta) \label{bclapseorb} \, . \eea
Convenient ans\"{a}tze for $\rho$ and $\chi^\mu$ must be introduced in
practice to cope with these conditions.

\medskip

{\it b) Warped coordinate system}. An alternative choice consists
in adopting an evolution vector $t^\mu$ such that its boundary
value on each horizon coincides with $t_o^\mu$, but adapts itself
to $L^\mu$ at a typical distance $\delta$ from them (in this
sense, the distance between the black holes provides a natural
length scale in the binary problem). Hence, this vector $t^\mu$
interpolates between $t_o^\mu$ and $L^\mu$, {\it warping} the
coordinate system to better accommodate the physical situation in
each of the considered spatial regions (note that the vector
$L^\mu$ follows the translational motion, whereas $t^\mu_o$ is
adapted to the intrinsic rotation on the horizon). Of course such
a coordinate system can remain regular only for a finite amount of
time (typically one orbital period).

In practical terms, this coordinate system is defined by the outer
boundary conditions (\ref{outerbc}), without primes in $\beta^i$
and $\alpha$, and the inner boundary conditions (\ref{bcshift})
[on ${\cal S}$], (\ref{bcpsi}), (\ref{trazatheta}), and
(\ref{bclapse}) on the constrained and free data. Moreover, the
function $tr_{_{\cal S}} \dot{\tilde{\gamma}}$ does not have to
vanish on ${\cal S}$, therefore helping to ensure the positivity
of $\Psi$, even though it must be negligible at a distance of
order $\delta$. Likewise, $\partial_t K$ and the {\it radial}
components $\partial_t\tilde{\gamma}_{rj}$ become roughly zero at
a distance $\delta$ of each horizon. Thus, in this coordinate
system, the conditions on the horizons are easier to impose, there
is no need to worry about the factor $\rho$, and one gains control
over the positivity of $\Psi$.

\medskip

Once the time derivative part of the free data has been fixed,
either in the corotating or in the warped coordinate system, one
would have to consider the rest of the free data. The choice of
the conformal metric must be consistent with restriction
(\ref{NEHtotal2}) [and (\ref{trazathetaorb}) in corotating
coordinates] and subject to the constraint ${\rm
det}(\tilde{\gamma}_{ij})=f$. The adequate determination of the
physical content of $\tilde{\gamma}_{ij}$ goes beyond the limited
scope of this paper and must be addressed by means of a proper
analysis of the stationary regime for the evolution equations
(\ref{evoleq}).

As for gauge fixing, the Dirac gauge in Ref. \cite{BGGN03} appears
to be a quite natural choice for the spatial one in the CTS
setting, whereas the boundary condition (\ref{bcpsi}) might be
viewed to suggest a maximal slicing ($K=0$) for the temporal gauge
in order to improve the control on the positivity of $\Psi$.
However, this latter gauge is not compatible with coordinates of
Painlev\'e-Gullstrand or Kerr-Schild type, which are actually
appropriate for the shift boundary condition (\ref{bcshift}). We
do not here subscribe to a particular fixation of the gauge,
allowing an optimal adaptation to each case considered.

\section{Comments on previous approaches}

\subsection{Cook's 2002 proposal}

Inner boundary conditions for the elliptic Eqs. (\ref{psieq}),
(\ref{shifteq}), and (\ref{lapseeq}) in the quasi-circular regime
of a binary black hole system were presented in Ref. \cite{Co01}.
The scheme proposed in that work starts by imposing on each
excised sphere ${\cal S}$ the presence of a Killing horizon,
together with an {\it apparent horizon} condition,
$\theta_{(\tilde l)}\,\stackrel{\Delta}{=}\,0$, where ${\tilde
l}^\mu$ is defined by Eq.~(\ref{scaling}). Denoting by $\zeta^\mu
= \alpha n^\mu + \beta^\nu s_\nu s^\mu$ the component of $t^\mu$
orthogonal to ${\cal S}$, the following quasi-equilibrium
conditions were imposed:
\begin{enumerate}
\item The inner boundary ${\cal S}$ {\it remains}
an apparent horizon: $ {\cal L}_\zeta \theta_{(\tilde l)}
\,\stackrel{\Delta}{=}\,0 $.
\item The expansion $\theta_{(\tilde{k})}$
associated with the ingoing null vector $\tilde{k}^\mu = (n^\mu -
s^\mu)/2$ does not {\it change} in time: $ {\cal L}_\zeta
\theta_{(\tilde{k})}\,\stackrel{\Delta}{=}\,0 $ .
\end{enumerate}
These conditions are enforced under the {\it approximation},
motivated by the stationary case, that the shear $\sigma_{(\tilde
l)}$ associated with the outgoing null vector vanishes.

Before we actually compare the resulting boundary conditions (or
rather some ellaboration of them; see the last part of this
section) with those of Sec. IV, we make some general remarks on
the involved quasi-equilibrium conditions.

Under the vanishing shear approximation, the condition $ {\cal
L}_\zeta \theta_{(\tilde l)}\,\stackrel{\Delta}{=}\,0$ leads to
$\beta^\nu s_\nu\,\stackrel{\Delta}{=}\,\alpha$, thus making
$\zeta^\mu$ a null vector parallel to $\tilde{l}^\mu$. In
particular, this implies that the underlying coordinate system is
stationary with respect to the horizon. Therefore, this condition
is either redundant with the vanishing shear approximation (via
Raychaudhuri equation) or must be considered as a gauge choice,
and not as an actual quasi-equilibrium condition.

More generally, in Ref. \cite{Co01} the conceptual status of the
vanishing shear hypothesis is not clearly stated and an explicit
prescription for imposing it in terms of the initial data, such as
Eqs. (\ref{trazatheta}) or (\ref{trazathetaorb}), is missing. The
IH analysis shows that the vanishing of the shear is the key
quasi-equilibrium condition: it guarantees that the world-tube of
apparent horizons is a null hypersurface. More explicitly, if
$\sigma_{(\tilde l)}\,\,\stackrel{\Delta}{=}\,\,0$ is not taken as
a quasi-equilibrium {\it characterization} but only as an
approximation that might occasionally fail, the vector $\zeta^\mu$
is no longer necessarily null. As a consequence ${\cal L}_\zeta
\theta_{(\tilde l)}\,\stackrel{\Delta}{=}\,0$ would not really be
a quasi-equilibrium condition (for instance, ${\cal L}_\zeta
\theta_{(\tilde l)}$ vanishes also for dynamical horizons
\cite{AK03}, where $\zeta^\mu$ is spacelike).

Hence, as already pointed out in Refs. \cite{DKSS02,AK04}, the
approach of Ref. \cite{Co01} is very close in spirit to that
encoded in the IH formalism; in fact, if the approximation of
vanishing shear is eventually satisfied, a NEH is actually
constructed. However, its quasi-equilibrium conditions can be
refined \footnote{We have obviated the discussion of Ref.
\cite{Co01} about the way to enforce the horizon to remain in the
same coordinate location, a discussion that can also be
simplified.} (see Ref. \cite{Co03} and Subsec. V.A). By contrast,
a virtue of our approach, fully based on the IH scheme, is a clear
identification and understanding of the physical and mathematical
hypotheses that characterize the horizon quasi-equilibrium.

The need to clarify, from a conceptual point of view, the
quasi-equilibrium hypotheses in Ref. \cite{Co01} can be
illustrated as follows. In Ref. \cite{YCSB04}, the boundary
conditions derived in Ref. \cite{Co01}, identified as IH
conditions, are disregarded as technically too complicated. They
are then substituted by a heuristic set of conditions, involving
in particular $0=\partial_t \ln\sqrt{\gamma}= D_i \beta^i - \alpha
K$. This condition, which is equivalent to $\partial_t \Psi=0$,
turns out to be the NEH condition on the conformal factor in the
corotating physical regime [make $\eta^\mu=0$ in Eq.
(\ref{bcpsiorb})], under the quasi-equilibrium bulk condition
$\partial_t \tilde{\gamma}_{ij}=0$ assumed in Ref. \cite{YCSB04}.
Besides, the Kerr-Schild data (motivating the boundary values for
the shift in that reference) are consistent with Eq.
(\ref{bcshiftorb}). At the end of the day, we find that the
heuristic choice turns out to be one which is truly in the spirit
of the IH scheme. Let us nonetheless mention that these boundary
conditions are not imposed on the horizon itself, but in its
interior.

\subsection{Addendum}

After the first submission of this work, a paper by Cook and
Pfeiffer appeared \cite{CP04} which provides a refinement of the
discussion and proposals made by Cook in Ref. \cite{Co01}. We now
comment on the relation between our approach and the
quasi-equilibrium and boundary conditions proposed in Ref.
\cite{CP04} (and in \cite{Co01}), in order to facilitate the
comparison of our results with those of that reference.

1. {\it Quasi-equilibrium conditions}. Quasi-equilibrium is
characterized in Ref. \cite{CP04} by the geometrical conditions
$\theta_{(\tilde l)}\,\stackrel{\Delta}{=}\, \sigma_{(\tilde
l)}\,\stackrel{\Delta}{=}\,0$, which are exactly those required to
construct a NEH horizon, as discussed in Subsec. IV.B. Since no
other condition is imposed (the requirement ${\cal
L}_{\zeta}\theta_{(\tilde{k})}\,\stackrel{\Delta}{=}\,0$ of Ref.
\cite{Co01} is dropped), the analysis remains at the level of a
NEH in the IH hierarchy, whereas our approach explores the WIH
structure.

2. {\it Condition on $\Psi$}. In Refs. \cite{Co01, CP04}, this
boundary condition follows from the requirement of vanishing
expansion $\theta_{(\tilde l)}$ for an apparent horizon. It is
therefore essentially equivalent to Eqs. (\ref{bcpsi}) and
(\ref{bcpsiorb}). However, the mathematical expression derived
from $\theta_{(\tilde l)}\stackrel{\Delta}{=}0$ adopts different
forms [see also Eq. (\ref{expansionspat}) in Appendix A].

3. {\it Condition on $\beta^i$}. In {\it corotating} coordinates,
the boundary condition (79) of Ref. \cite{Co01} essentially
coincides with our Eq. (\ref{bcshiftorb}). A crucial refinement is
introduced in Ref. \cite{CP04} by actually imposing that the shear
vanish: the projection of the shift on ${\cal S}$ [our vector
$-\eta^\mu$ in Eq. (\ref{eta})] must be a conformal symmetry of
$\tilde{q}_{ab}$. This is equivalent to our condition
(\ref{trazathetaorb}) (see also Appendix A).

The main difference between both approaches is our demand of an
azimuthal symmetry $\varphi^\mu$ for the metric $q_{ab}$, namely
Eq. (\ref{NEHtotal2}). On the one hand, this makes conditions in
Ref. \cite{CP04} more general than ours but, on the other hand,
thanks to this symmetry we are able to introduce a definite,
intrinsic spinning angular velocity $\Omega_\Delta$ which,
together with $\Omega_{orb}$, permits to analyze the rotational
regime of the system (corotational, irrotational or general case).

In addition, the availability of $\Omega_\Delta$ naturally leads
us to consider Eq. (\ref{bcshiftorb}) as a boundary condition on
the shift. As a consequence, Eq. (\ref{trazathetaorb}) becomes a
constraint on the free data $\tilde{\gamma}_{ab}$ and $\partial_t
\tilde{\gamma}_{ab}$, rather than providing a boundary condition
for $\eta^\mu$ as in Ref. \cite{CP04}.

4. {\it Condition on $\alpha$}. The analysis of a WIH carried out
in Subsec. IV.C shows that the initial boundary value for the
lapse is basically free. This conclusion is also reached in Ref.
\cite{CP04} after a numerical study. It is worth discussing the
relation between the proposals that have been made for the choice
of this boundary value. Condition $ {\cal L}_\zeta \theta_{(\tilde
{k})}\,\stackrel{\Delta}{=}\,0 $ in Ref. \cite{Co01} can be
written as
\begin{equation} \left.\left[s^i D_i \alpha - s^i s^j
K_{ij}\alpha\right]\right|_{\cal S}= -\left.\frac{\tilde{\cal
D}\alpha}{\theta_{(\tilde{k})}}\right|_{\cal S}
 \ , \label{alphacook} \end{equation}
with $\tilde{\cal D}$ defined in Eq. (85) of Ref. \cite{Co01}. Our
requirement (\ref{bclapse}) and Eq. (\ref{alphacook}) are
simultaneously satisfied only if $\tilde{\cal
D}\alpha=-\theta_{(\tilde{k})}\kappa_{_{Kerr}}$ on ${\cal S}$.
This is a nontrivial identity, so that both conditions are
generally different.

Insight on their relation is provided by Ref. \cite{ABL01b}, where
the freedom in the construction of a WIH structure $[l^\mu]$ is
fixed by imposing that ${\cal
L}_{(l)}\theta_{(k)}\,\stackrel{\Delta}{=}\,0$ (with $k^{\mu}l_\mu
= -1$), once it is assumed that a certain operator ${\bf M}$ which
acts on ${\cal S}$ has a trivial kernel [see Eq. (4.8) of Ref.
\cite{ABL01b} for the definition of ${\bf M}$ and note its close
connection with $\tilde{\cal D}$]. This analysis can be applied to
study the possible degeneracy of condition (\ref{alphacook}) in
terms of the invertibility of ${\bf M}$ if, in addition, it is
satisfied that $\kappa_{(\tilde{l})}$ is constant. If that is the
case, employing Eq. (\ref{kaltil}) one can check that conditions
(\ref{bclapse}) and (\ref{alphacook}) coincide only if the lapse
is constant on the boundary. More details on this issue will
appear elsewhere.

Notice that the choice of representative made in a WIH class via
the lapse boundary condition (\ref{bclapse}) determines the
initial lapse once $\gamma_{ij}$ and $K_{ij}$ are given. In this
sense, the condition for the lapse is not problematic by itself.
However, our full set of boundary conditions for $\alpha$, $\Psi$
and $\beta^i$, together with the choice of free data
$(\tilde{\gamma}_{ij},\partial_t \tilde{\gamma}^{ij}, K,
\partial_t K)$, may not be sufficient to single out a unique
solution to the Initial Data Problem. In fact, this degeneracy
seems to occur when our boundary conditions are implemented in the
spherically symmetric, time-independent case if one uses a maximal
slicing and a flat conformal metric\footnote{We thank G.B Cook for
pointing out this fact that also happens with the boundary
conditions of Ref. \cite{Co01}.}. Nonetheless, the presence of
this degeneracy may depend on the actual choice of initial free
data (e.g., isotropic coordinates in the commented example). For
each specific choice, it is generally only after a numerical study
that one may decide whether a degeneracy exists.

\section{Conclusions}

In this work we have explicitly shown how the IH formalism
provides a rationale for some aspects of the numerical
construction of initial data for a space-time containing a black
hole in local quasi-equilibrium, with special emphasis in the
binary case.

The IH framework sheds light into the justification and
implications of already existing quasi-equilibrium sets of
conditions for the analysis of this problem. The hierarchical
structure of the IH formalism permits a control on the hypotheses
that arise at each of the considered steps.

Adopting the IH approach fully, we have derived a set of boundary
conditions on each black hole horizon for solving the elliptic
equations obtained in a CTS scheme, deduced from the constraints
and the trace of the evolution part in Einstein equations.

In a first step, the NEH condition characterizing
quasi-equilibrium
($\Theta_{ab}\,\stackrel{\Delta}{=}\,0\Leftrightarrow
\theta_{(l)}\,\stackrel{\Delta}{=}\,{\sigma_{(l)}}_{ab}
\,\stackrel{\Delta}{=}\,0$), together with the choice of spatial
coordinates stationary with respect to the horizon, provides
boundary conditions for the shift [see Eqs. (\ref{bcshift}) and
(\ref{bcshiftorb})] and the conformal factor [see Eqs.
(\ref{bcpsi}) and (\ref{bcpsiorb})]. These conditions are
basically equivalent to those of Ref. \cite{Co01} (at least in the
recently refined form presented in Ref. \cite{CP04}). In a second
step, the requirement for a WIH-compatible slicing
($\kappa_{(l)}\stackrel{\Delta}{=}{\rm constant}\neq 0$) leads to
the evolution equation (\ref{alphaevol}) for the lapse on
$\Delta$, leaving the choice of its initial value essentially
free. Once this point has been acknowledged, we have tentatively
suggested a specific boundary condition for the lapse in Eq.
(\ref{bclapse}), obtained by fixing the freedom which is available
in the construction of a WIH structure. In addition to these
boundary conditions, the NEH requirement entails that the free
data on the initial slice fulfill, on the horizon, the constraints
(\ref{trazatheta}) [or (\ref{trazathetaorb}) in corotating
coordinates] and (\ref{NEHtotal2}) (assuming an axially symmetric
horizon; see Ref. \cite{fin} otherwise). Similarly, Eq.
(\ref{alphaevol}) could alternatively be seen as a constraint on
the free data if one decided to fix the lapse on $\Delta$.

These inner boundary conditions and constraints are sufficient
conditions for constructing a black hole in instantaneous
quasi-equilibrium. However, in order to obtain black holes in
quasi-equilibrium during a finite evolution time rather than just
instantaneously (as required in the quasi-circular binary black
hole problem), these conditions must be complemented with
appropriate free data that encode the desired dynamical behavior.

\acknowledgments

The authors are greatly grateful to A. Ashtekar, S. Bonazzola, S.
Dain, B. Krishnan, and J. Novak for enlightening conversations and
comments. They are also very thankful to C. Barcel\'o, G.B. Cook,
J. Lewandowski, J.M. Mart\'{\i}n-Garc\'{\i}a, H. Pfeiffer, J. Pullin, and K.
Uryu for helpful discussions. J.L.J. was funded by a postdoctoral
fellowship from the Spanish MEC and G.A.M.M. was supported in part
by funds provided by the Spanish MEC projects BFM2002-04031-C02
and BFM2001-0213.

\appendix
\section{Some Technical details}

In this appendix we explain some calculations and formulas
employed in the main text.

{\it a) Proof of Eq.} (\ref{pregeodesic}). If $\Delta$ is defined
as the hypersurface $r={\rm constant}$, its normal $l^\mu$ takes
the form $l_\mu=\lambda\nabla_\mu r$ for certain function
$\lambda$. Then $\nabla_{[\mu}l_{\nu]} = \lambda^{-1}
l_{[\nu}\nabla_{\mu]}\lambda$. Contracting with $l^\mu$, we find
\bea l^\mu\nabla_\mu l_\nu=l^\mu\nabla_{\mu}\ln \lambda \; l_\nu
\, .\eea Eq. (\ref{pregeodesic}) follows by identifying
$\kappa_{(l)}\equiv l^\mu \nabla_{\mu}\ln\lambda$.

{\it b) Proof of Eq.} (\ref{omega}). Choosing the normalization of
the outgoing and ingoing null vectors $l^\mu$ and $k^\mu$ on
$\Delta$ so that $k^\mu l_\mu=-1$, the induced (degenerate) metric
on $\Delta$ can be written as \bea q_{\mu\nu} = g_{\mu\nu}+k_\mu
l_\nu+ l_\mu k_\nu=g_{\mu\nu}+n_\mu n_\nu - s_\mu s_\nu
\label{qkl} \ . \eea Expressing ${\cal L}_l q_{\mu\nu}$ in terms
of the connection $\nabla_\mu$ and using Eq. (\ref{qkl}), we get
for the second fundamental form (\ref{sec}) the formula: \bea
\Theta_{\alpha\beta}= {q^\mu}_\alpha {q^\nu}_\beta \nabla_\mu
l_\nu\ . \label{thetanabla} \eea Expanding now ${q^\mu}_\alpha
{q^\nu}_\beta$ and recalling Eq. (\ref{pregeodesic}), we find \bea
\nabla_\alpha l_\beta &=& \Theta_{\alpha\beta}- l_\alpha k^\nu
\nabla_\nu l_\beta \nonumber\\ &-& (k^\nu\nabla_\alpha l_\nu +
l_\alpha k^\mu k^\nu \nabla_\mu l_ \nu)\, l_\beta \ . \eea Since,
on a NEH, one has $\Theta_{\alpha\beta}\stackrel{\Delta}{=} 0$
(see the text), contraction with a vector $v^\mu$ tangent to
$\Delta$ (so that $v^\mu l_\mu\stackrel{\Delta}{=}0$) leads to Eq.
(\ref{omega}) after defining \bea \omega_\alpha
\stackrel{\Delta}{=}- (k^\nu\nabla_\alpha l_\nu + l_\alpha k^\mu
k^\nu\nabla_\mu l_ \nu)= -{P^\mu}_\alpha k^\nu\nabla_\mu l_\nu \,
.\eea

{\it c) Proof of Eq.} (\ref{WIHcharac}). To demonstrate this
equation starting from condition (\ref{Lw}), the key remark is the
proportionaliy between the exterior derivative of $\omega_\mu$ and
the volume 2-form on the sphere $S^2\simeq{\cal S}$, \bea {\rm d}
\omega \propto \sqrt{q}\; d^2q \ . \eea This is a nontrivial
result that follows from the definition of NEH, and we refer the
reader to Ref. \cite{AFK00} for its proof. As a consequence, ${\rm
d} \omega$ {\it lives} on ${\cal S}$ and its contraction with the
null normal of $\Delta$ vanishes: $l^\mu
\nabla_{[\mu}\omega_{\nu]}\stackrel{\Delta}{=}0$. Therefore, using
the Cartan identity, \bea 0\stackrel{\Delta}{=} {\cal L}_l \omega_
\nu =l^\nu \nabla_{[\nu}\omega_{\mu]} +\nabla_\mu\left(l^\nu
\omega_\nu \right)\stackrel{\Delta}{=} \nabla_\mu \kappa_{(l)}\, .
\eea

{\it d) General expression of $\Theta_{ab}$}. In order to enforce
the condition $\Theta_{ab}\stackrel{\Delta}{=}0$, we have made use
in the text of a coordinate system that is stationary with respect
to the horizon, so that $l^\mu=t^\mu + W^\mu$ with $W^\mu$ tangent
to ${\cal S}$ ($W^\mu$ is the black hole surface velocity
introduced by Damour \cite{Damou78,Damou79}). Then, we have
considered several specific choices for $t^\mu$ on $\Delta$,
namely, the vector $t_o^\mu$ singled out in Eq. (\ref{to}) for the
{\it warped} coordinate system, or the quasi-Killing vector
$L^\mu$ for {\it corotating} coordinates. For the sake of
completeness, we now provide the general expression of the second
fundamental form on $\Delta$ for arbitrary vectors $t^\mu$ and
$W^\mu$.

From the definition of $\Theta_{\mu\nu}$ and Eq. (\ref{qkl}) we
find \bea \Theta_{ab}&=&\frac{1}{2} \left({\cal L}_t q_{ab} +
^2\!\!D_a W_b + ^2\!\!D_b W_a\right ) \nonumber \\ &=&\frac{1}{2}
\left({\cal L}_t q_{ab} + {\cal L}_W q_{ab} \right) \ .\eea Here,
$^2\!D_a$ denotes the connection compatible with the metric
$q_{ab}$ induced on the sphere ${\cal S}$ and $W_a \equiv q_{ab}
W^b$. Note that Eqs. (\ref{Ltq}) and (\ref{condgen}) follow
straightforwardly from this when one imposes $W^\mu$ to be the
$q_{ab}$-isometry $\Omega_{_\Delta}\varphi^\mu$.

The conformal decomposition $q_{ab}=\Psi^4\tilde{q}_{ab}$ leads to
\bea \Theta_{ab}=\frac{\Psi^4}{2}\left[\theta_{(l)}\tilde{q}_{ab}
+ {\cal L}_t\tilde{q}_{ab} -\frac{1}{2}\left({\cal L}_t \ln
\tilde{q}\right)\tilde{q}_{ab} \right. \nonumber \\ \left.+
^2\!\tilde{D}_a {\tilde W}_b+ ^2\!\!\tilde{D}_b {\tilde W}_a
-^2\!\!\tilde{D}_c W^c\,\tilde{q}_{ab}\right] \ , \label{thetagen}
\eea and
\begin{equation}
\theta_{(l)} \!=\! \frac{1}{2}{\cal L}_t \ln \tilde{q} +4{\cal
L}_t \ln \Psi \!+ ^2\!\!\tilde{D}_a W^a \!\!+ 4 W^a\;
{}^2\!\tilde{D}_a \ln \Psi , \label{expansiongen}
\end{equation}
where $^2\!\tilde{D}_a$ is the connection compatible with
$\tilde{q}_{ab}$ and ${\tilde W}_a \equiv {\tilde q}_{ab} W^b$. In
particular, with the notation $\eta^\mu$ for $W^\mu$ and choosing
$t^\mu$ to be the quasi-killing vector $L^\mu$ (so that
$\partial_t\tilde{q}_{ab}=0$), the vanishing of the expansion in
Eq. (\ref{expansiongen}) leads to condition (\ref{bcpsiorb})
[after substituting Eq. (\ref{psidot}) for $\partial_t \Psi$]. On
the other hand, from Eq. (\ref{thetagen}), the traceless part of
$\Theta_{ab}\stackrel{\Delta}{=}0$ is equivalent to condition
(\ref{trazathetaorb}). It is then clear from Eq. (\ref{thetagen})
that in {\it corotating} coordinates the vector $\eta^\mu$ must be
a conformal symmetry generator of $\tilde{q}_{ab}$.

Finally, a more general expression for $\Theta_{\alpha\beta}$ can
be obtained if we do not assume a coordinate system stationary
with respect to $\Delta$. Substituting $l^\mu=\alpha(n^\mu+s^\mu)$
in Eq. (\ref{thetanabla}) and expanding the derivative we find
\bea \Theta_{\alpha\beta} = \alpha\left( D_\mu s_\nu -
K_{\mu\nu}\right){q^\mu}_\alpha {q^\nu}_ \beta \ \ . \eea Taking
the trace, the standard expression for $\theta_{(l)}$ follows,
\bea \theta_{(l)}= \alpha(D_i s^i+K_{ij} s^i s^j - K) \ . \eea
Hence, after a conformal decomposition, \bea \theta_{(l)} &=&
\left(4\tilde{s}^i\tilde{D}_i \ln \Psi +\tilde{D}_i\tilde{s}^i +
\Psi^{-2}K_{ij} \tilde{s}^i\tilde{s}^j -\Psi^{2} K\right)
\nonumber
\\ &&\times
\alpha \Psi^{-2} \, , \label{expansionspat} \eea where
$\tilde{s}^i=\Psi^{2} s^i$.


\begin{thebibliography}{99}
\bibitem{fin} If the hypothesis of axial symmetry on the horizon
were not adequate, e.g. because of tidal distortions owing to the
black hole companion, NEH conditions would still apply in a
slightly modified form. For instance, conditions
(\ref{condgenpart}) would be replaced by Eq. (\ref{condgen}), with
the quantity $J_\Delta$ being still well-defined (see Refs.
\cite{ABL01a,DKSS02}) though devoided of its angular momentum
interpretation.
\end{thebibliography}
\end{document}